\documentclass[conference]{IEEEtran}

\usepackage{cite}
\usepackage{url}
\usepackage{amsmath,amssymb,amsfonts}
\usepackage{algorithmic}
\usepackage{graphicx}
\usepackage{lipsum}
\usepackage{hhline}
\usepackage{tabularx}
\usepackage{xcolor}
\usepackage{multirow}
\usepackage{subcaption}
\usepackage{dblfloatfix}    
\usepackage[font=small]{caption}

\graphicspath{{graphics/}}
\usepackage{algorithm}
\usepackage{algorithmic}

\makeatletter
\def\url@leostyle{%
	\@ifundefined{selectfont}{\def\UrlFont{\sf}}{\def\UrlFont{\small\ttfamily}}}
\makeatother
\DeclareMathOperator*{\argmax}{arg\,max} 

\DeclareMathOperator*{\softmax}{soft\,max} 
\DeclareMathOperator*{\sortmax}{sort\,max}

\def\BibTeX{{\rm B\kern-.05em{\sc i\kern-.025em b}\kern-.08em
		T\kern-.1667em\lower.7ex\hbox{E}\kern-.125emX}}
\begin{document}
	
	\title{Learning-based Remote Radio Head Selection and Localization in Distributed Antenna System\\
		\thanks{}
	}
	
	\author{\IEEEauthorblockN{1\textsuperscript{st} Given Name Surname}
		\IEEEauthorblockA{\textit{dept. name of organization (of Aff.)} \\
			\textit{name of organization (of Aff.)}\\
			City, Country \\
			email address or ORCID}
		\and
		\IEEEauthorblockN{2\textsuperscript{nd} Given Name Surname}
		\IEEEauthorblockA{\textit{dept. name of organization (of Aff.)} \\
			\textit{name of organization (of Aff.)}\\
			City, Country \\
			email address or ORCID}
		\and
		\IEEEauthorblockN{3\textsuperscript{rd} Given Name Surname}
		\IEEEauthorblockA{\textit{dept. name of organization (of Aff.)} \\
			\textit{name of organization (of Aff.)}\\
			City, Country \\
			email address or ORCID}
	}
	
	\author{
		\IEEEauthorblockN{Artan Salihu$^\dagger$$^\ddagger$, Stefan Schwarz$^\dagger$$^\ddagger$ and Markus Rupp$^\dagger$}
		\IEEEauthorblockA{$^\dagger$ Institute of Telecommunications, Technische Universit{\"a}t (TU) Wien\\
			$^\ddagger$ Christian Doppler Laboratory for Dependable Wireless Connectivity for the Society in Motion \\
			Email: \{artan.salihu,stefan.schwarz,markus.rupp\}@tuwien.ac.at\\
		}
	}
	
	\maketitle
	\begin{abstract}
		In this work, we consider estimating user positions in a spatially distributed antenna system (DAS) from the uplink channel state information (CSI). However, with the increased number of remote radio heads (RRHs), collecting CSI at a central unit (CU) can significantly increase the fronthaul overhead and computational complexity of the CU. This problem can be mitigated by selecting a subset of RRHs. Thus, we present a deep learning-based approach to select a subset of RRHs for wireless localization. We employ an RRH selection layer that is jointly trained with the rest of the network and learn the model parameters as well as the set of selected RRHs. We show that the selection strategy comes at a relatively small cost of localization performance. Nonetheless, by comparison to a trivial approach based on the maximization of the channel gain, we show that the proposed method leads to significant performance gains in a propagation environment dominated by non-line-of-sight.
	\end{abstract}
	
	\vspace{-0.27cm}
	\section{Introduction}
	The ever-growing demand for location-enabled applications has caused a surge of research interest to enhance the accuracy and dependability of wireless positioning methods for both indoor and outdoor environments. Furthermore, future mobile communication systems can exploit location information for various tasks, including better capacity planning, optimization, and dynamic allocation of resources. This is particularly beneficial as we pursue network densification, namely adding more antennas and base stations (e.g., to increase the spectrum spatial reuse).
	
	There exist different strategies to exploit the channel properties of the transmitter to determine its position, such as received signal strength (RSS), angle-of-arrival (AOA), and time of arrival (TOA) \cite{zane2020performance,wen2019survey}. Lately, the deployment of massive multiple-input multiple-output (MIMO) technology in the fifth generation (5G) networks has encouraged active research in machine learning (ML) and, in particular, deep learning for wireless positioning \cite{9348191, salihu2020low, sun2019fingerprint}. In this case, due to large antenna arrays at the massive MIMO base station (BS), one can leverage from a considerable amount of high-dimensional channel state information (CSI) to train a model with examples from known locations. The model then utilizes the CSI of the unknown transmitter to infer its position.
	
	In order to provide a more uniform quality of service for users in the region of interest (ROI), distributed antenna systems (DAS) are being advocated. In this case, a large number of spatially distributed antennas are utilized by means of remote radio heads (RHHs) in order to extend the BS antenna ports \cite{schwarz2018remote}. Due to the better spatial diversity, DAS can be favorable for wireless positioning too. However, the main drawback of DAS is that it requires an expensive infrastructure to connect RRHs with a central unit (CU). Additionally, collecting signal information from a large number of RRHs further increases the computational burden at the CU. Hence, a selection method that utilizes CSI only from the ``best'' $M$ out of the $N$ RRHs at a small expense of localization performance is desired. The problem of interest is, in fact, a combinatorial optimization that is generally NP-hard.
	
	Regardless of the numerous benefits of DAS in wireless communications, the investigation regarding its positioning capabilities is limited. The works in \cite{savic2015fingerprinting,prasad2018machine} make use of RSS and machine learning (ML) based on Gaussian processes (GP), while \cite{qiu2020cooperative,wei2020joint} investigate other conventional out-of-the-box ML techniques and exploit different channel information. None of the former works considers any selection approach. In contrast, we present a deep learning method with the capability of RRH subset selection strategy, user localization and uncertainty estimation.
	
	We structure the remaining of this paper as follows. In Section \ref{systemModel}, we describe the system model considered for this work. Then, in Section \ref{uePosUncertainty}, we provide details on the RRH selection strategy and user location estimation. In Section \ref{experiments}, we evaluate the performance in DAS. In this section, we also analyze the effectiveness of the proposed selection strategy concerning the localization error. Finally, in Section \ref{conclusion}, we draw our conclusions.
	
	\textit{Notation:} A matrix is denoted by $\mathbf{Y}$, a vector by $\mathbf{y}$ and a scalar by $y$. The element of a matrix $\mathbf{Y}$ is $y_{m,n}$ and the Euclidean norm of vectors is $\| \cdot \|$. The cardinality of a set as well as the absolute value of a scalar are $| \cdot |$. A categorical variable is denoted by $\verb*|Cat|(\cdot)$, and a length $N$ binary row of a matrix as $\verb*|one_hot|_{N}(\cdot)$.
	\section{System Model}\label{systemModel}
\begin{figure}[!t]
		\centering
		{%
			\includegraphics[width=0.75\linewidth]{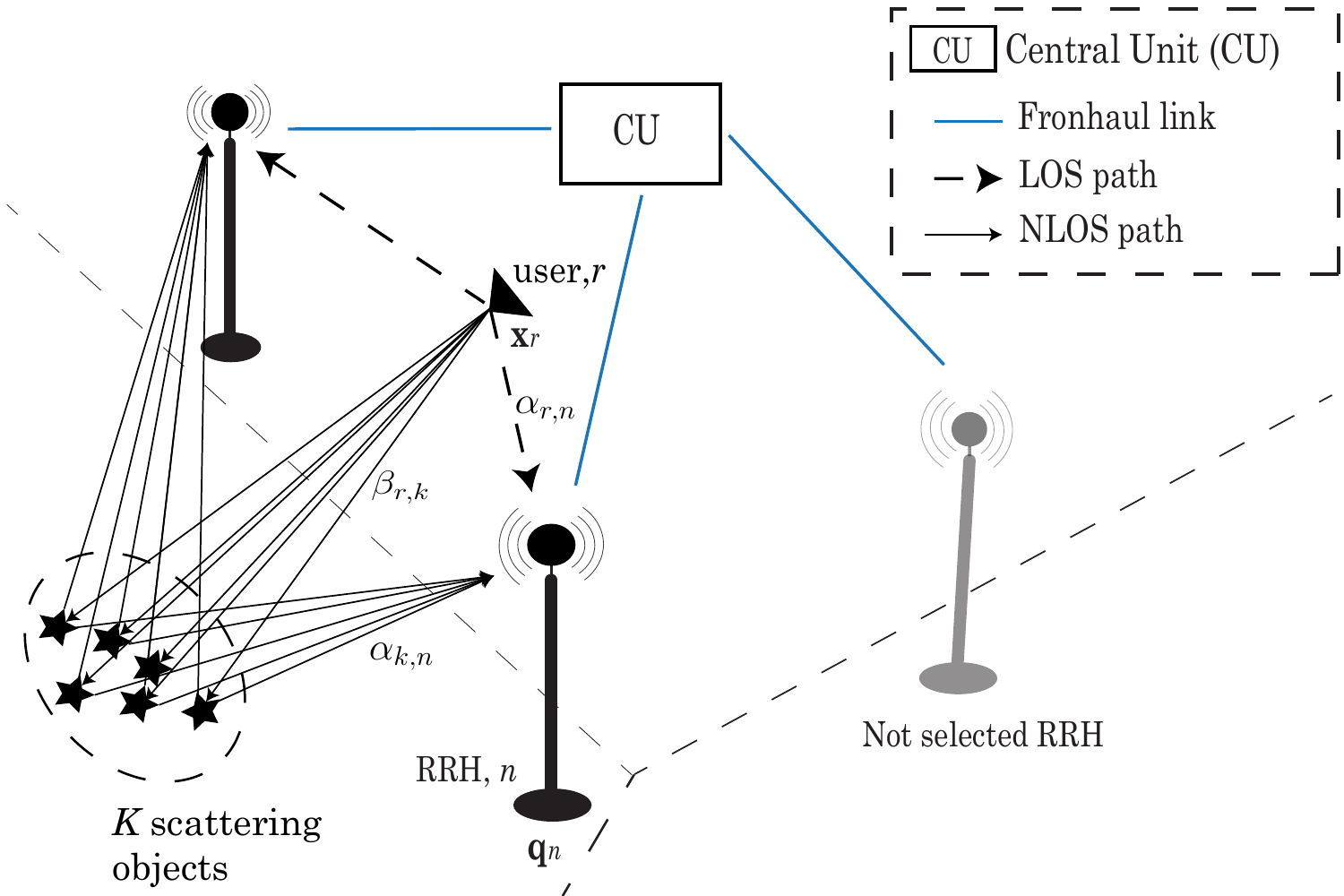}}	
		\caption{Illustration of the system model considered for RRH selection and user location estimation in a distributed antenna system (DAS).}
		\label{fig:System_Model} \vspace{-0.4cm}
	\end{figure}

	We consider a distributed massive MIMO system, as shown in  Fig. \ref{fig:System_Model}. We assume $N$ spatially distributed and single-antenna remote radio heads (RRHs) at positions $\mathbf{q}_{n} \in \mathbb{R}^{2}$ with $n \in \mathcal{N}$, where $\mathcal{N}$ denotes the set of RRH indices and $\left| \mathcal{N} \right| = N$. Assuming mutually orthogonal	pilot sequences across the users, each RRH receives the wireless signal from $R$ single-antenna transmitters placed at positions $\mathbf{x}_{r} \in \mathbb{R}^{2}$ with $r \in \mathcal{R}$, where $\mathcal{R}$ denotes the set of user indices and $\left| \mathcal{R} \right| = R$. Further, we consider $K$ scattering objects in the ROI at respective positions, $\mathbf{p}_{k}\in\mathbb{R}^{2}$ with $k \in \mathcal{K}$, where $\mathcal{K}$ denotes the set of scattering object indices and $\left| \mathcal{K} \right| = K$. 
	
	We assume that all RRHs are connected via high-speed fronthaul links to the central unit (CU), i.e., the delay between the RRHs and the CU is negligible. Initially, each RRH obtains signal information from $R$ individual locations during the training phase and forwards it to the CU. The role of the CU is to process the per-user channel estimates and learn a subset of RRHs. Then, throughout the operations phase, the role of the CU becomes to perform positioning inference of the unknown user transmitter with the channel information gathered from the subset of RRHs, which are learned during the preceding phase.
	The channel between the $r-$th user location and the $n-$th RRH is given by a spatially consistent and scenario-specific channel model \cite{pratschner2019spatially} \vspace{-0.1cm}
	\begin{equation}
		h_{r, n}=\alpha_{r, n} + \sum_{k = 1}^K \beta_{r, k} \alpha_{k, n} \quad,
	\end{equation}
	where $\alpha_{r, n}$ is the direct line-of-sight (LOS) path, i.e., it describes the propagation without involving any scattering. The sum term describes the signal over $K$ other paths that include a scattering event due to the scattering objects placed in the ROI. As in \cite{pratschner2019spatially}, path coefficients $\alpha_{i, j}$ and $\beta_{i, j}$ are obtained as: \vspace{-0.0cm}
	\begin{equation}
		\begin{array}{l}
			\alpha_{i, j}=\frac{\lambda}{4 \pi\left\|\mathbf{v}_{i}-\mathbf{v}_{j}\right\|} e^{i \frac{2 \pi}{\lambda}\left\|\mathbf{v}_{i}-\mathbf{v}_{j}\right\|} , \\ \\
			
			\beta_{i, j}=\frac{\delta_{j}}{\sqrt{4 \pi}\left\|\mathbf{v}_{\ell}-\mathbf{v}_{j}\right\|} e^{i \frac{2 \pi}{\lambda}\left\|\mathbf{v}_{i}-\mathbf{v}_{j}\right\|} . \label{eqn:scattering_coefficients}
		\end{array}
	\end{equation}
	The coefficients in (\ref{eqn:scattering_coefficients}) describe the wave propagation characteristics between any two elements $i$ and $j$ at their respective positions $\mathbf{v}_i$ and $\mathbf{v}_{j}$ in the ROI. The scattering event $\delta_{j}$ captures the electromagnetic properties of the scattering objects, and $\lambda$ denotes the carrier wavelength. We assume that scattering objects impose a random phase-shift, $\varphi_{j}$, of the signal and have an amplitude gain of $\gamma_{j}=\left|\delta_{j}\right|$. The amplitude gain is equal to the square root of the bi-static radar cross section (RCS) of the scattering object. As shown in [11], it can be related to the Rician K-factor of the effective multipath channel and, therefore determines the relative strength of LOS and NLOS components. Consequently, we are able to model a more realistic propagation environment that distinguishes from the extreme LOS and non-line-of-sight (NLOS) cases in order to better understand the performance limits of our methods in terms of the localization error. The resulting channel model, after substituting the scattering coefficients (\ref{eqn:scattering_coefficients}) and assuming $\gamma_j = \gamma, \forall j$, is given as \vspace{-0.2cm}
	\begin{equation}
		h_{r, n}=\frac{\lambda}{4 \pi\left\|\mathbf{x}_{r}-\mathbf{q}_{n}\right\|} e^{i \frac{2 \pi}{\lambda}\left\|\mathbf{x}_{r}-\mathbf{q}_{n}\right\|}+\frac{\lambda \gamma}{(4 \pi)^{\frac{3}{2}}} \sum_{k=1}^K u\left(\mathbf{p}_{k}\right).\label{eqn:channel_coef_sum}\vspace{-0.1cm}
	\end{equation}
	The individual sum terms are further expressed as
	\begin{equation}
		\begin{aligned}
			u\left(\mathbf{p}_{k}\right)=\left(\left\|\mathbf{x}_{r}-\mathbf{p}_{k}\right\|\left\|\mathbf{p}_{k}-\mathbf{q}_{n}\right\|\right)^{-1} & \\
			& \hspace{-2cm}\times e^{i\left(\varphi_{k}+\frac{2 \pi}{\lambda}\left(\left\|\mathbf{x}_{r}-\mathbf{p}_{k}\right\|+\left\|\mathbf{p}_{k}-\mathbf{q}_{n}\right\|\right)\right)}. \label{eqn:sum_term}
		\end{aligned}
	\end{equation}
	We maintain a minimum distance between any transmit antenna location and RRH, $\| \mathbf{x}_{n} - \mathbf{q}_{n}\| > 0 $, as well as between any scattering object and RRH, $\| \mathbf{q}_{n} - \mathbf{p}_{n} \| > 0$. Finally, for each user $r$, the CU forms a $N \times 1$ channel vector $\mathbf{h}_r$, i.e., \vspace{-0.0cm}
	\begin{equation}
		\mathbf{h}_{r}=\left[h_{r,1}, h_{r,2}, \ldots, h_{r,N}\right]^{T} \in \mathbb{C}^{N\times 1} . \label{eqn:channel_vector}
	\end{equation} 
	In this work, we rely on channel strength information for localization, $\widetilde{\mathbf{h}}_{r} = |\mathbf{h}_{r}|$.
	\vspace{0.3cm}
	
	\section{Localization with RRHs selection strategy}\label{uePosUncertainty}
	
	\subsection{Problem Formulation}
	Our goal is to reduce the computational complexity at the CU by determining which RRHs should be selected while providing accurate and reliable localization performance. The selection process is fixed and performed once per ROI. The selected RRHs are used for all the test users over a period of time, until the next scheduled selection process occurs. 

\begin{figure*}[!t]
		\includegraphics[width=1.0 \textwidth,height=4.5cm]{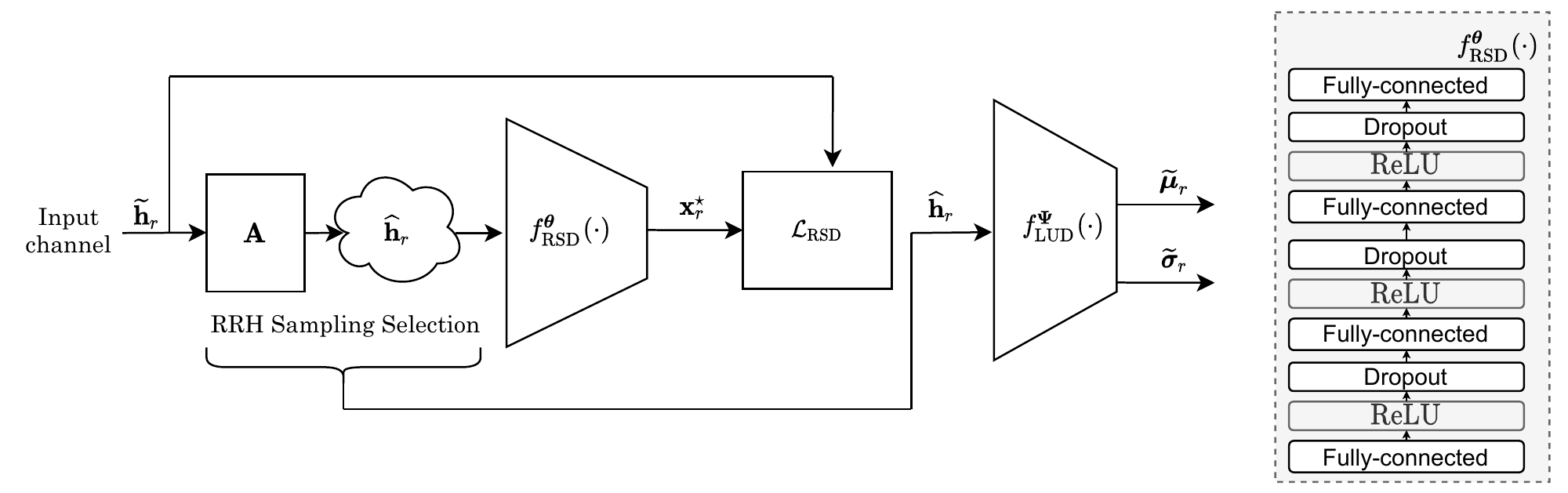}
		\caption{RRH selection DNN ($\text{RSD}$) with a sampling and selection layer and localization with uncertainty estimation DNN ($\text{LUD}$).} \label{fig:RSD_and_LUD} \vspace{-0.5cm} 
	\end{figure*}

	We approach this problem in two steps. First, we describe the developed end-to-end and DNN-based RRH selection method. Then, using the subset of RRHs, we elaborate how to utilize another DNN to estimate the final position of the transmitter and the uncertainty associated with it. An overview of the RRH selection and localization is shown in Fig. \ref{fig:RSD_and_LUD}.
	
	Specifically, we seek to learn a subset $\mathcal{M} \subseteq \mathcal{N}$ of spatially distributed RRH indices of $| \mathcal{M} | = M$, as well as, a channel to location mapping function $f_{\textrm{RSD}}^{\boldsymbol{\theta}}:\mathbb{R}^{M} \mapsto \mathbb{R}^{2}$, where $\boldsymbol{\theta}$ denotes the parameters (weights) of the first DNN model, which we will refer to as RRH selection DNN, or \textrm{RSD}. Thus, we seek to minimize the localization error between the estimated $\mathbf{x}_{r}^{\star} = f_{\textrm{RSD}}^{\boldsymbol{\theta}}(\widehat{\mathbf{h}}_{r})$ and the given location information $\mathbf{x}_{r}$. Formally, our goal becomes to learn the optimal $\mathcal{M}^{\star}$ and $\boldsymbol{\theta}^{\star}$ such that  \vspace{-0.0cm}
	\begin{equation}
		\begin{aligned}
			\left\{  \mathcal{M}^{\star}, \boldsymbol{\theta}^{\star} \right\} &=\underset{\mathcal{M}, \boldsymbol{\theta}}{\arg \min }  \, \mathcal{L}_{\textrm{RSD}}(\mathbf{x}_{r}, \mathbf{\widehat{{h}}}_{r}, \boldsymbol{\theta}) \\
			&= \underset{\mathcal{M}, \boldsymbol{\theta}}{\arg \min }  \, \mathbb{E}\left[\left\|f_{\textrm{RSD}}^{\boldsymbol{\theta}}(\mathbf{\widehat{{h}}}_{r})-\mathbf{x}_{r} \right\|^{2}\right].\vspace{-0.3cm}
		\end{aligned}\label{eqn:loss_RSD}
	\end{equation}
	In (\ref{eqn:loss_RSD}), $\mathbf{\widehat{{h}}}_{r}$ contains CSI from $M$ selected RRHs. It can be written as\vspace{-0.0cm}
	\begin{equation}
		\begin{aligned}
			\mathbf{\widehat{h}}_{r} 
			= \mathbf{A} \mathbf{\widetilde{h}}_{r} \;, \label{eqn:rrh_compression_formulation}\vspace{-0.1cm}
		\end{aligned}
	\end{equation} 
	where $\mathbf{A} \in \{0,1\}^{M \times N}$ represents a binary selection matrix with $m$ selection-rows $\mathbf{a}_{m}$, i.e.,  $a_{m,n} = 1$ if $n \in \mathcal{M}$. 
	
	\subsection{RRH Selection}
	The RRH selection in (\ref{eqn:loss_RSD}), is a combinatorial optimization problem over the discrete set $\mathcal{M} \subseteq \mathcal{N}$ of RRHs. This implies that the number of RRH subsets grows exponentially with $N$ and $M$. In addition, implementation difficulties arise due to the inability to backpropagate over the discrete variables. 
	To circumvent these difficulties, we rely on a \textit{reparameterization} approach \cite{kingma2015variational}: instead of directly optimizing over $\mathbf{a}_{m}$, we relax it by a learnable $N-$dimensional vector $\boldsymbol{\phi}_m \in \mathbb{R}_{+}^{N}$. We do so by introducing an RRH selection layer, trained together with the rest of the network, that utilizes \emph{concrete} variables \cite{balin2019concrete,maddison2016concrete}, based on a Gumbel distribution \cite{gumbel1954statistical}. In this case, the RRH selection layer controls the intensity of relaxation of the elements in the binary matrix $\mathbf{A}$, using the layer temperature $\tau \in (0,\infty)$.
	
	Formally, we begin by defining $z_{m}$ to be a categorical variable over the domain $\{ 1, 2, \ldots, N \}$, which can be represented by a vector of class probabilities $\boldsymbol{\pi}_{m} \in \mathbb{R}^{N}$, i.e., 
	\begin{equation}
		z_{m} \sim \verb*|Cat|(\boldsymbol{\pi}_{m}, N),\label{eqn:Categorical_Variable}
	\end{equation}
	where the class probabilities are related to vector $\boldsymbol{\phi}_{m}$ as 
	\begin{equation}
		\pi_{m,n}=\frac{\exp \phi_{m,n}}{\sum_{i=1}^{N} \exp \phi_{m,i}}.
	\end{equation}
	
	Using the Gumbel-Max trick \cite{gumbel1954statistical,NIPS2014_309fee4e}, we can sample from (\ref{eqn:Categorical_Variable}) by independently perturbing $\phi_{m,n}$ with i.i.d. Gumbel noise, $g_{m,n} = \texttt{Gumbel}(0,1)$. Specifically, we draw $N$ samples from a uniform distribution and form $\mathbf{u}_{m} \sim \mathcal{U}(0,1)$. Then, we apply the $\texttt{Gumbel}$ inverse CDF, $F_{G}^{-}(\mathbf{u}_m)$. In more detail,
	\begin{equation}
		\begin{aligned}
			\boldsymbol{g}_{m} &= F_{G}^{-}(\mathbf{u}_{m}) = -\log(-\log(\mathbf{u}_{m})) \label{eqn:Gumbel_Distribution}
		\end{aligned}
	\end{equation} 
	and
	\begin{equation}
		\tilde{z}_{m} = \argmax_{n \in \{ 1,2,\ldots, N \}} \{ \phi_{m,n} + g_{m,n}\}.\label{eqn:Cat_argmax}
	\end{equation}
	The $\argmax$ of the respective perturbed probabilities returns a sample from the original categorical distribution, $\tilde{z}_{m} \stackrel{\text{d}}{=} z_{m}$.	
	%
	Finally, during the forward pass, we can convert $\tilde{z}_{m}$ into a binary vector $\mathbf{a}_{m}$, such that $\|\mathbf{a}_{m}\| = 1$ and we write
	\begin{equation}
		\begin{aligned}
			\mathbf{a}_{m} &= \verb*|one_hot|_{N} \{ \tilde{z}_{m} \} \\
			&= \verb*|one_hot|_{N}\{ \argmax_{n} \{ \phi_{m,n} + g_{m,n} \}\}. 
		\end{aligned} \label{eqn:Gumbel_one_hot} \vspace{-0.2cm}
	\end{equation}
	However, in order to allow gradient backpropagation during the training phase, we need to relax the non-differentiable $\argmax(\cdot)$ operation. To do so, we rely on Gumbel-Softmax \cite{maddison2016concrete} and approximate the $\argmax(\cdot)$ in (\ref{eqn:Cat_argmax}) with a continuous soft operation, $\softmax_{\tau} (\cdot)$.
	Then, $\mathbf{a}_m$ is replaced by
	\begin{equation}
		\begin{aligned}
			\mathbf{a}_{m} = \lim _{\tau \rightarrow 0} \frac{\exp \left\{ \tau^{-1} \left(\boldsymbol{\phi}_{m}+\boldsymbol{g}_{m}\right) \right\}}{\sum_{n=1}^{N} \exp \left\{ \tau^{-1} \left(\phi_{m,n}+g_{m,n}\right) \right\}} . \vspace{-0.2cm}
		\end{aligned} \label{eqn:Gumbel_Softmax}
	\end{equation} 
	As $\tau \rightarrow 0 $, (\ref{eqn:Gumbel_Softmax}) smoothly approaches the $\argmax$, and each of the neurons in the RRH sampling and selection layer outputs one of the selected RRHs. During the training over a total number of $T$ epochs, the value of $\tau$ decreases exponentially, and for the $t-$th epoch, it is given by
	\begin{equation}
		\tau_{t} = \tau_{0}\left(\tau_{\text{e}} / \tau_{0}\right)^{t / T}, \label{eqn:Temperature} \vspace{-0.2cm}
	\end{equation}
	where $\tau_{\text{0}}$ and $\tau_{\text{e}}$ are the user-defined start and end temperature values, respectively. This enables the end-to-end network to explore the combinations of RRHs in the beginning and move from linearly combining to selecting a particular RRH by the end of the training. The training process repeats until each unit in the selection layer only selects one specific RRH, $m \in \mathcal{M}$. 
	
	\subsection{DNN Learning and Operation}
	During the test time, i.e., the RRH selection process and localization, we drop the stochasticity of the network and replace the continuous $\softmax_{\tau}(\cdot)$ by a discrete $\argmax (\cdot)$. The training as well operation phase of the $\textrm{RSD}$ $\forall{m} \in \mathcal{M}$ are further summarized in Algorithm \ref{alg:RRH_training_Alg} and Algorithm \ref{alg:RRH_testing_Alg}, respectively.  
	
	\begin{algorithm}[H]
		\hspace*{\algorithmicindent} \textbf{Given:} A training dataset $\mathcal{D} = \{\mathbf{h}_r, \mathbf{x}_r\}_{r=1}^{R}$, number of RRHs to select $M$, epochs $T$, learning rate $\alpha$, final temperature $\tau_{\text{e}}$, start temperature $\tau_{0}$ \vspace{0.1cm}\\ 
		\hspace*{\algorithmicindent} Build $\textrm{RSD}$,  $f_{\textrm{RSD}}^{\boldsymbol{\theta}}(\cdot)$, and initialize $\boldsymbol{\theta}$.\\ \vspace{-0.3cm}
		\begin{algorithmic}[1]
			\FOR{$m \in \{ 1,2, \ldots, M \} $}
			\STATE Initialize a $N-$ dimensional vector $\boldsymbol{\phi}_m \in \mathbb{R}_{+}^{N}$.
			\ENDFOR
			\FOR{$ t \in \{ 1, 2, \ldots, T \}$}
			\STATE Compute $\tau$ according to (\ref{eqn:Temperature})
			\FOR{$m \in \{ 1,2, \ldots, M \}$}
			\STATE Compute $\mathbf{a}_{m}$ using (\ref{eqn:Gumbel_one_hot})
			\STATE Set $\widehat{h}_{r,m} = \mathbf{a}_m \widetilde{\mathbf{h}}_{r}$			
			\ENDFOR	
			\STATE Compute the gradients of $\mathcal{L}_{\textrm{RSD}}$ w.r.t. $\boldsymbol{\theta}$ and each $\boldsymbol{\phi}_{m}$.
			\STATE Update the parameters $\boldsymbol{\theta} \leftarrow \boldsymbol{\theta} - \alpha \nabla_{\boldsymbol{\theta}} \mathcal{L}_{\textrm{RSD}}$ and  $\boldsymbol{\phi} \leftarrow \boldsymbol{\phi} - \alpha \nabla_{\boldsymbol{\phi}} \mathcal{L}_{\textrm{RSD}}$
			\ENDFOR
		\end{algorithmic}
		\hspace*{\algorithmicindent} \textbf{Return:} Trained $f_{\textrm{RSD}}^{\theta}(\cdot)$ and learned parameters $\boldsymbol{\theta}$ and $\boldsymbol{\phi}$.
		\caption{Training RSD}
		\label{alg:RRH_training_Alg}
	\end{algorithm}
	\vspace{-0.2cm}
	\begin{algorithm}[H]
		\hspace*{\algorithmicindent} \textbf{Given:} A test channel $\widetilde{\mathbf{h}}_{r}$ and trained parameters $\boldsymbol{\phi}$\\ \vspace{-0.4cm}
		\begin{algorithmic}[1]
			\FOR{$ m\in \{ 1,2, \ldots, M \} $}
			\STATE Compute $\tilde{z}_{m} = \argmax_{n} \{ \phi_{m,n}\}$
			\STATE Set $\widehat{h}_{r,m} = \widetilde{h}_{r,\tilde{z}_{m}}$
			\ENDFOR
		\end{algorithmic}
		\hspace*{\algorithmicindent} \textbf{Return:} $\widehat{\mathbf{h}}_{r}$.
		\caption{Using RSD for RRH Selection}
		\label{alg:RRH_testing_Alg}
	\end{algorithm}
	
	To derive the final location estimate, the obtained $\mathbf{\widehat{{h}}}_{r}$ from $\text{RSD}$ is fed into the subsequent localization and uncertainty estimation DNN ($\textrm{LUD}$), which we recently proposed \cite{salihu2021towards}. Given $\mathbf{\widehat{{h}}}_{r}$, the second network is parameterized by $\boldsymbol{\Psi}$ and defined as $f_{\textrm{LUD}}^{\boldsymbol{\Psi}}:\mathbb{R}^{M} \mapsto \mathbb{R}^{4}$. This model, provides not only estimates of the position but also the uncertainty inherent in the samples and the model parameters $\boldsymbol{\Psi}$. Thus, the final location of the transmitter and its corresponding uncertainty are denoted by $\widehat{\boldsymbol{\mu}}_{r}$ and $\widehat{\boldsymbol{\sigma}}_{r}$, respectively. For details, see \cite{salihu2021towards}.
	
	\section{Experiments and Results}\label{experiments}
	In this section, we describe the parameter details for the investigated scenario, training details, and the results relating to the investigation of the proposed approach.\vspace{-0.0cm}
	\subsection{Simulation Parameters and Training Detail}\label{sec:simulation_parameters}
\begin{figure}[!h] 
		\centering
		{%
			\includegraphics[width=0.80\linewidth]{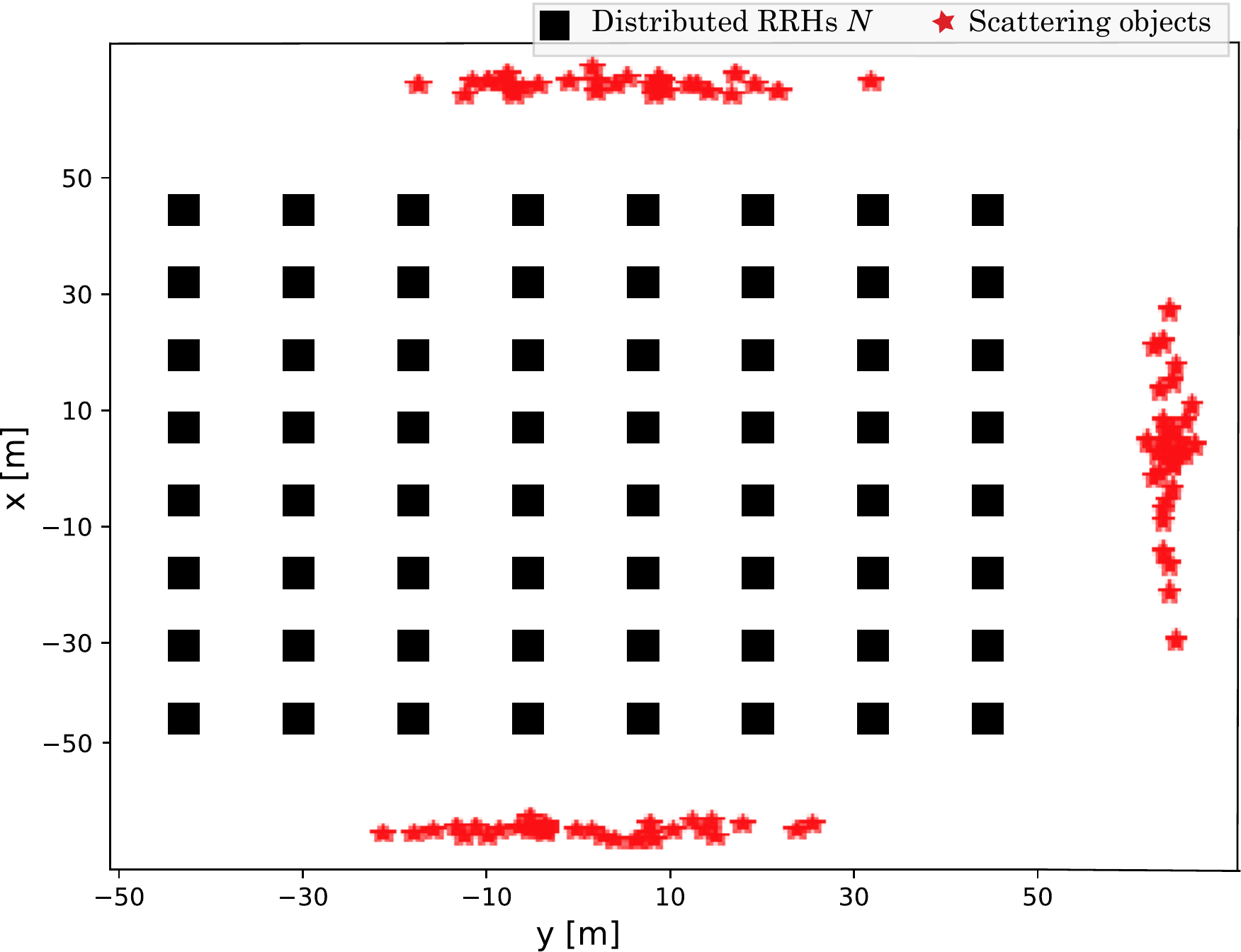}}	
		\caption{Simulation setup with $N$ spatially distributed RRHs.}
		\label{fig:Simulation_setup} \vspace{-0.4cm}
	\end{figure}

	We evaluate the proposed approach on a scenario where we consider $N \in \{ 16, 36, 64 \}$ RRHs distributed in a grid configuration within the ROI of $100\mathrm{m} \times 100\mathrm{m}$ range. An example of the simulation scenario with $N = 64$ is depicted the Fig. \ref{fig:Simulation_setup}. We place $K = 100$ scattering objects normally distributed in clusters $i \in \{1,2,3\}$ with  $\mathcal{N} (\boldsymbol{\mu}_{k}^{(i)}, \boldsymbol{\sigma}_{k}^{(i)})$, where $\boldsymbol{\mu}_{k}^{(1)} = [0, -60]$, $\boldsymbol{\mu}_{k}^{(2)} = [60, 0]$, $\boldsymbol{\mu}_{k}^{(3)} = [0, 60]$, $\boldsymbol{\sigma}_{k}^{(1)} = [100, 1]$, $\boldsymbol{\sigma}_{k}^{(2)} = [1, 100]$, and $\boldsymbol{\sigma}_{k}^{(3)} = [100, 1]$. The scattering coefficient is set as $\gamma = \{ 1.2, 3.0 \}$ to provide insights on the impact of NLOS components of the multipath channel. We uniformly sample from $R = 49,000$ locations in the ROI and we set the minimum distance $\| \mathbf{x}_{r} - \mathbf{q}_{n} \| > 0.5 \text{m}$.
	
	We adopt $\mathrm{ReLU}$ for the intermediate non-linear operations. The number of intermediate layers for $\textrm{RSD}$ is $B = 3$ with $350$ units. The number of units in the selection layer is $M$. For the presented experiments, a dropout rate of $0.2$ was selected. We train $\textrm{RSD}$ for $800$ epochs with $\tau_{e} = 0.1$ and $\tau_{0} = 10.0$. For training, we utilize the Adam \cite{ruder2016overview} solver, with a batch size of $512$ at a fixed learning rate of $10^{-3}$, and an early stopping if the validation loss does not improve for $80$ epochs. The parameters, $\boldsymbol{\theta}$ and $\boldsymbol{\phi}$, are initialized with a Glorot normal initializer \cite{glorot2010understanding}. The dataset is split into 0.8 and 0.2 for training and hold out testing, respectively. Performance is reported in terms of
	\begin{equation}
		\mathrm{RMSE}=\sqrt{\frac{\sum_{r=1}^{R_{\mathrm{test}}}\left\|\mathbf{x}_{r}-\widetilde{\boldsymbol{\mu}}_{r}\right\|^{2}}{R_{\mathrm{test}}} },
	\end{equation}
	where $\mathbf{x}_{r}$ is the actual position of the test location and, $\widetilde{\boldsymbol{\mu}}_{r}$ is the estimated location.
	
	\subsection{DAS Localization Accuracy}
\begin{figure}[!t] 
		\centering
		{%
			\includegraphics[width=0.90\linewidth]{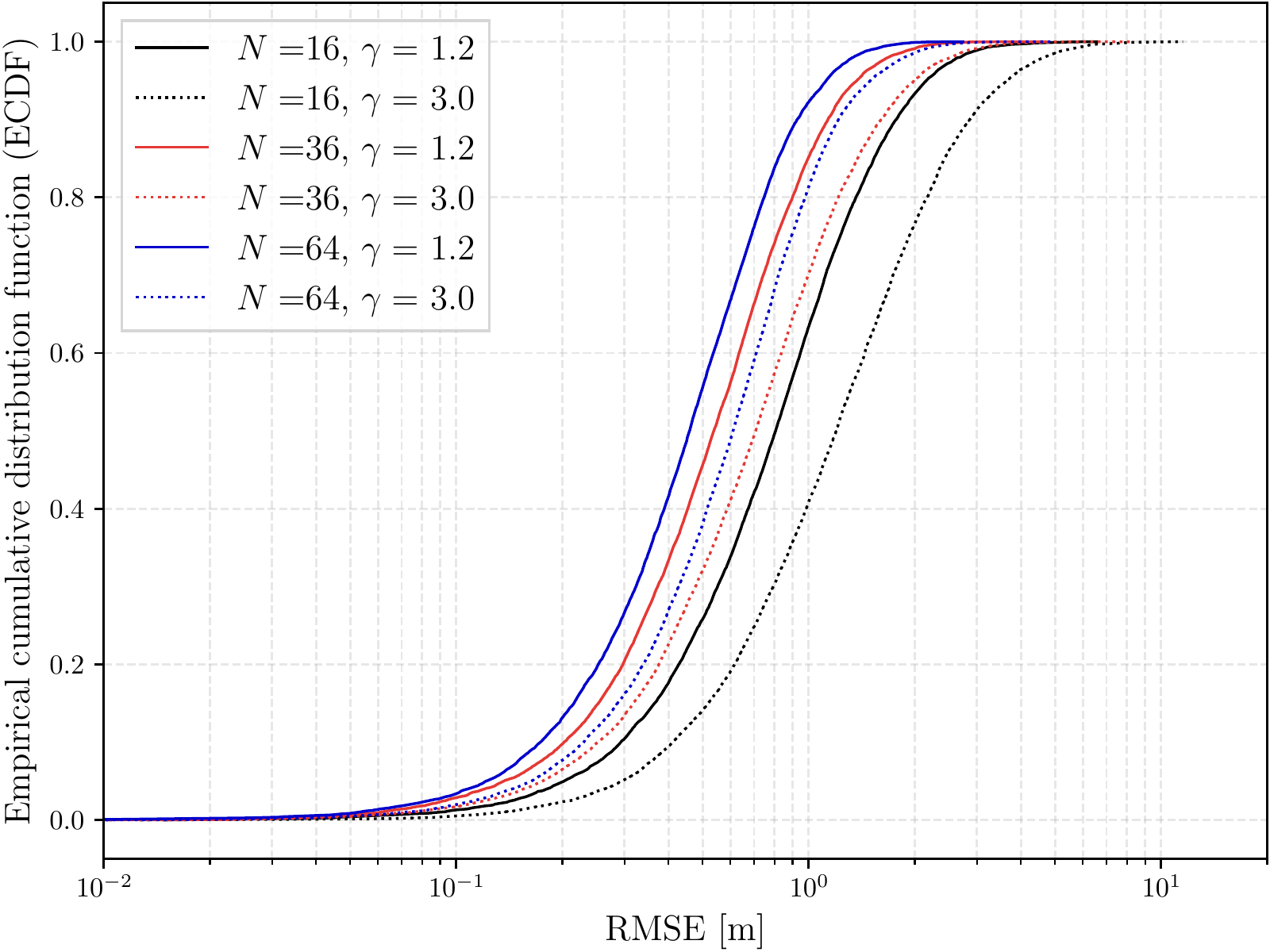}}
		\caption{Localization accuracy in DAS for different values of $N$ and $\gamma = \{1.2, 3.0\}$. The increase in RRHs improves the overall localization performance and understates the impact of NLOS.}
		\label{fig:ECDF_All} \vspace{-0.1cm}
	\end{figure}

	First, we investigate the localization accuracy in a DAS without any selection strategy. More specifically, we look into the impact of the number of RRHs as the NLOS components become stronger. This is shown in Fig. \ref{fig:ECDF_All}. We can observe that increasing $\gamma$, i.e., the value of scattering coefficients, the performance degrades in terms of the localization error for approximately $1 \mathrm{ m}$ at $90-$th percentile in case of $N = 16$. Naturally, increasing the number of RRHs improves the overall performance but can also compensate the impact of the dominant NLOS components, where the gap of the error at $90-$th percentile is approximately $30 \mathrm{ cm}$ for $N = 64$. 
	
\begin{figure}[!h]
		\centering
		\subfloat[ECDF - DAS vs centralized. \label{fig:ECDF_DAS-vs-Cent}]{%
			\includegraphics[width=0.95\linewidth]{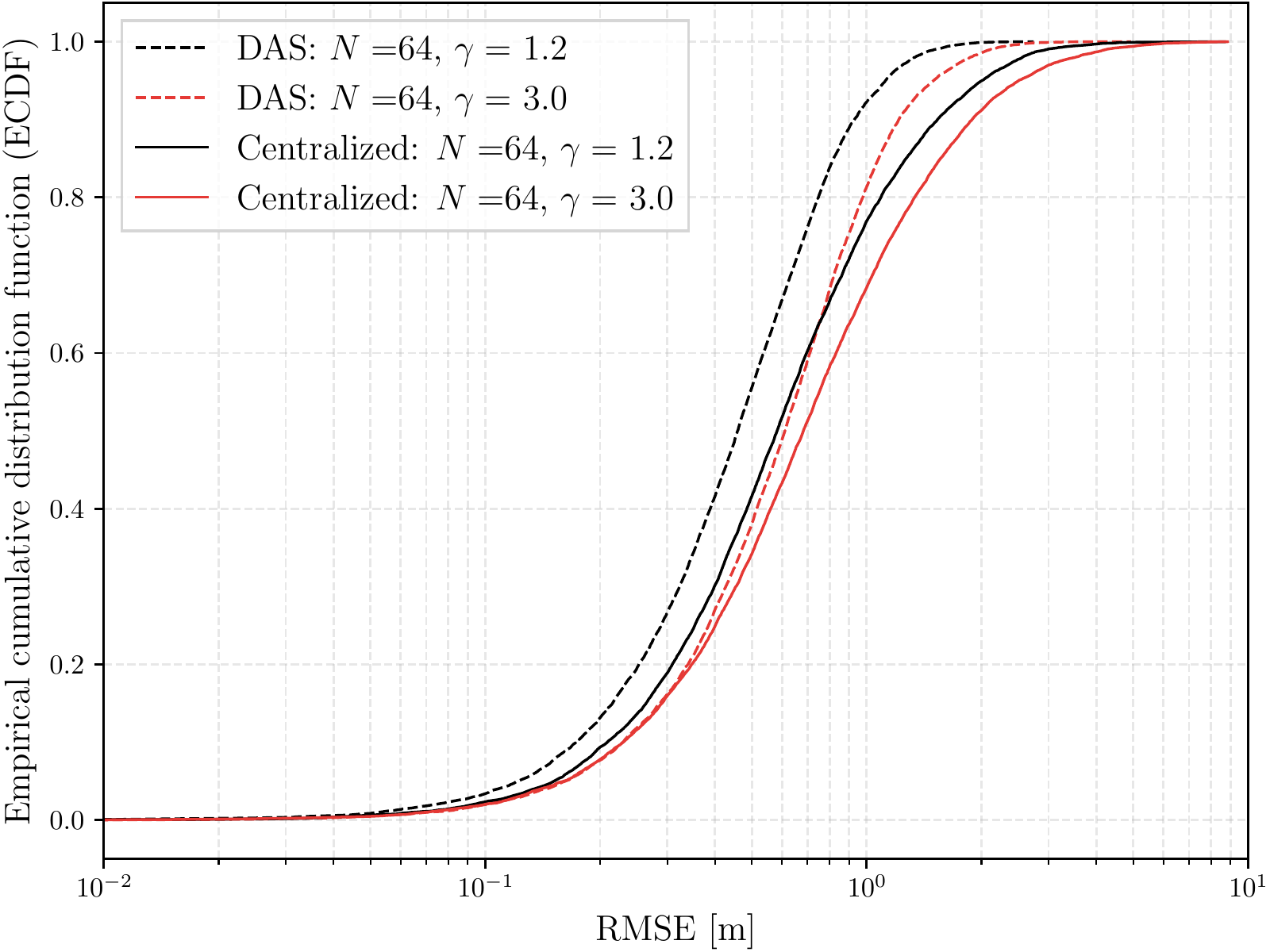}}
		\hfill
		\\
		\subfloat[Learned data uncertainty and respective $\mathrm{RMSE}$ for $N = 64$, and $\gamma=3.0$. Values are normalized for each case separately.  \label{fig:Uncertainty-Compared}]{%
			\includegraphics[width=0.90\linewidth]{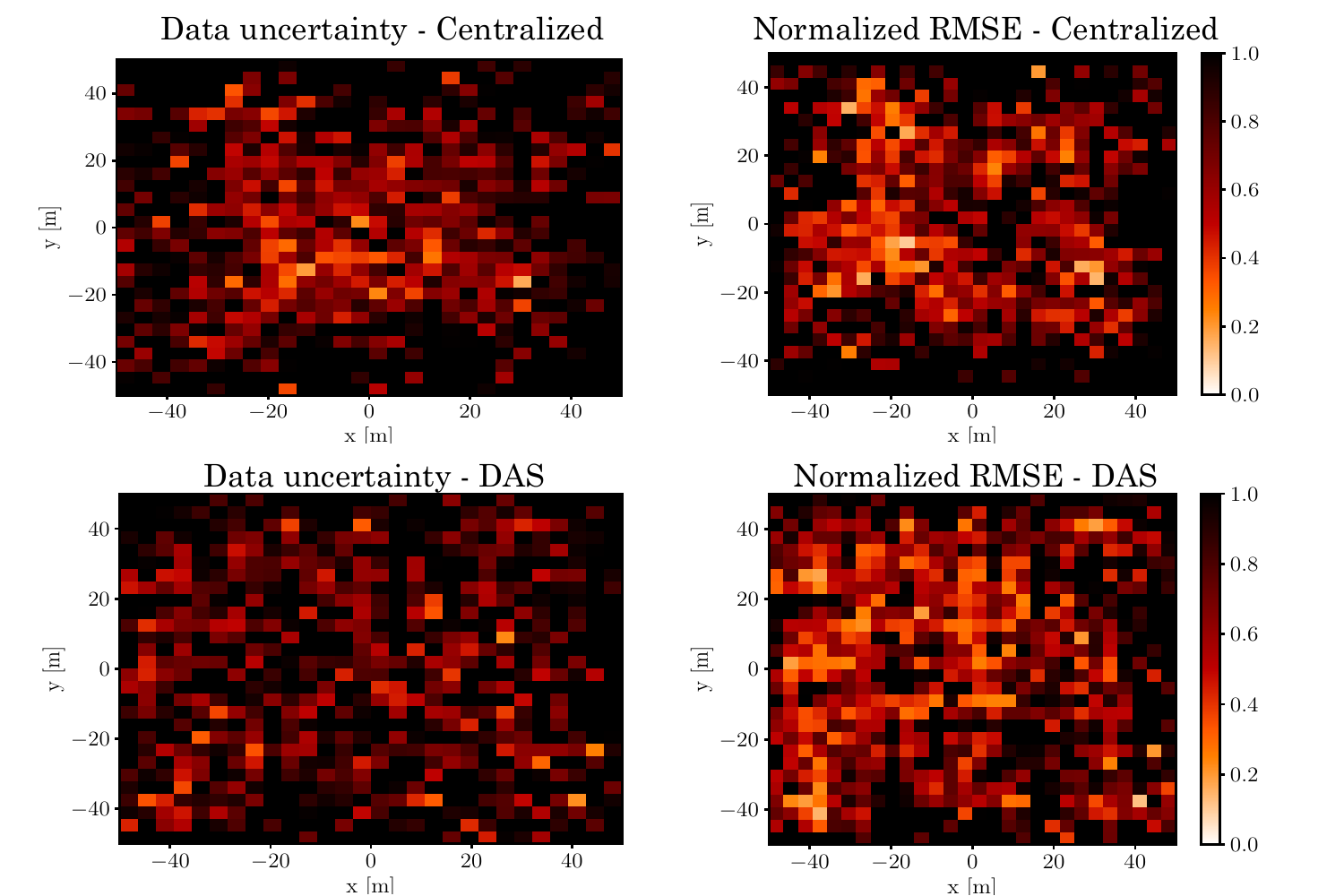}}\vspace{-0.0cm}	
		\caption{Comparison of DAS with centralized massive MIMO. DAS provides a more uniform error even for $\gamma = 3.0$.}
		\label{fig:DAS_vs_Cent_Uncertainty}\vspace{-0.0cm}
	\end{figure}

	Next, we compare the localization accuracy of a DAS with centralized massive MIMO for $N = 64$. In the case of centralized massive MIMO, we consider that antennas are aligned in a circular array with a diameter of $1.5 \mathrm{ m}$. Further, to include direction of arrival information and improve the granularity of obtained CSI, we handle the complex-valued coefficients as two independent real numbers, i.e., $\Re\left\{{h}_{r,n}\right\}, \Im\left\{{h}_{r,n}\right\}$, representing the real and imaginary components of $h_{r,n}$. Thus, the channel vector considered for comparison in centralized case is $\widehat{\mathbf{h}}_{r}\in \mathbb{R}^{2N}$. In Fig. \ref{fig:ECDF_DAS-vs-Cent}, we can observe that DAS improves the performance by more than $60 \mathrm{ cm}$ at $90-$th percentile. Moreover, in Fig. \ref{fig:Uncertainty-Compared}, we show that DAS provides a more uniform error across the ROI, which is fairly modeled by the learned data uncertainty too. Further increasing the number of antenna elements does not lower the error for the centralized massive MIMO.
	\subsection{RRH Selection Strategy}
\begin{figure}[!t] 
		\centering
		{%
			\includegraphics[width=0.90\linewidth]{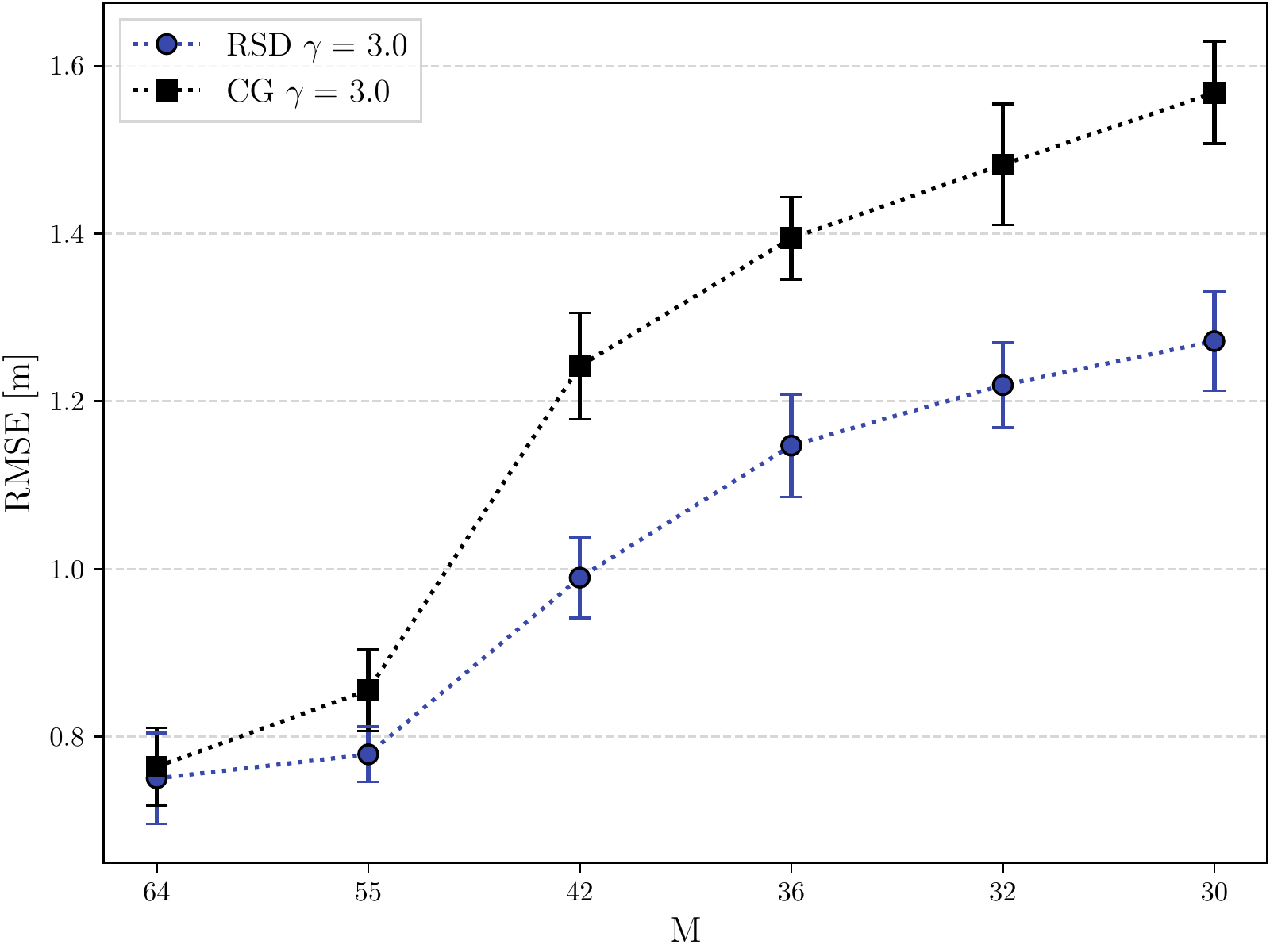}}	
		\caption{Localization accuracy for different $M$ and $\gamma = 3.0$. As the compression ratio increases, the performance gain is higher for RSD.}
		\label{fig:RSD_vs_CG} \vspace{0.0cm}
	\end{figure}

	To investigate the selection strategy, we evaluate the performance regarding the localization error for $M \in \{ 55, 42, 36, 32, 30 \}$. We compare the selection strategy to a low-complexity approach based on maximum channel gain (CG) at each RRH. In this case, we define $\sortmax_{N}(\cdot)$ operation, which provides $M$ indices of RRHs matching to the largest channel gain values in the descending order, i.e.,
	\begin{equation}
	\{z_1',...,z_M'\} =  \underset{n\in \mathcal{N}}{\sortmax_{M} } \, \sum_{r}\|h_{r,n} \|^{2} \,.
	\end{equation}
	Then, the CU selects $\widehat{h}_{r,m} =  \widetilde{h}_{r,z'_{m}}$, $\forall m$.
	In Fig. \ref{fig:RSD_vs_CG}, we show the averaged $\text{RMSE}$ over multiple realizations of $\textrm{LUD}$ and provide $95\%$ confidence interval for different $|\mathcal{M}|$. We see that the proposed selection strategy allows us to learn a subset of RRHs at a price of a small positioning performance loss. Compared to the straightforward CG approach, our method has a significant gain especially when $M \ll N$.
	
\begin{figure}[!h] 
		\centering
		{%
			\includegraphics[width=0.90\linewidth]{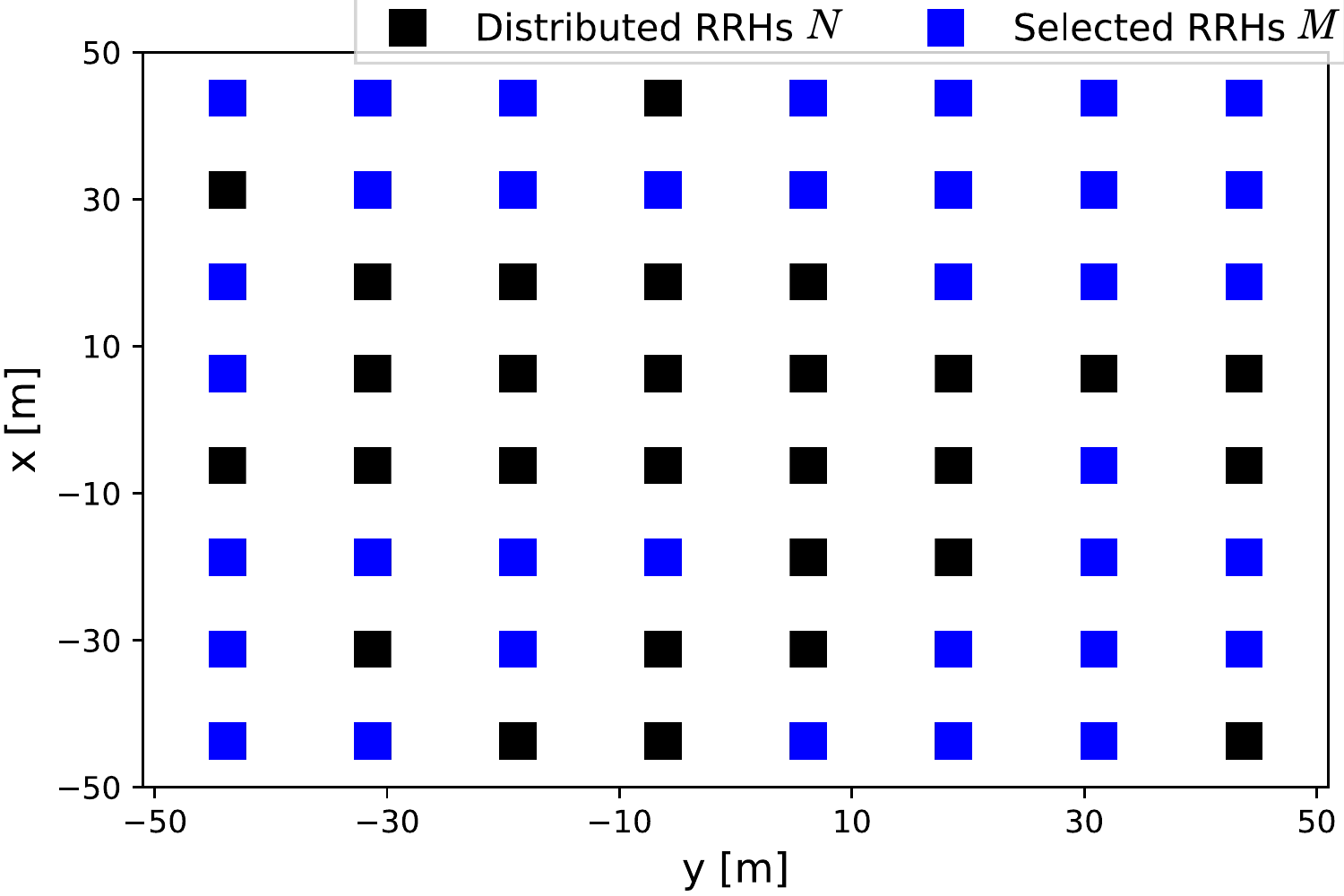}}
		\hfill
		\\
		\caption{Learned strategy for $M = 36$.}
		\label{fig:selected_dis_vs_grid}\vspace{-0.5cm}
	\end{figure}

	In  Fig. \ref{fig:selected_dis_vs_grid}, we show an example of the learned RRH selection pattern for $M = 36$  out of $N = 64$. To possibly understand how does the selection pattern change with $M$, next we investigated the probability of selecting $n \in \mathcal{N}$ as we vary the number of selected RRHs $|\mathcal{M}|$.
	
	\subsection{RRHs Revelant Positions}
\begin{figure}[!h] 
		\centering
		\subfloat[Frequency of selected top $M'=15$ RRHs \label{fig:frequency_bar_plot}]{%
			\includegraphics[width=0.85\linewidth]{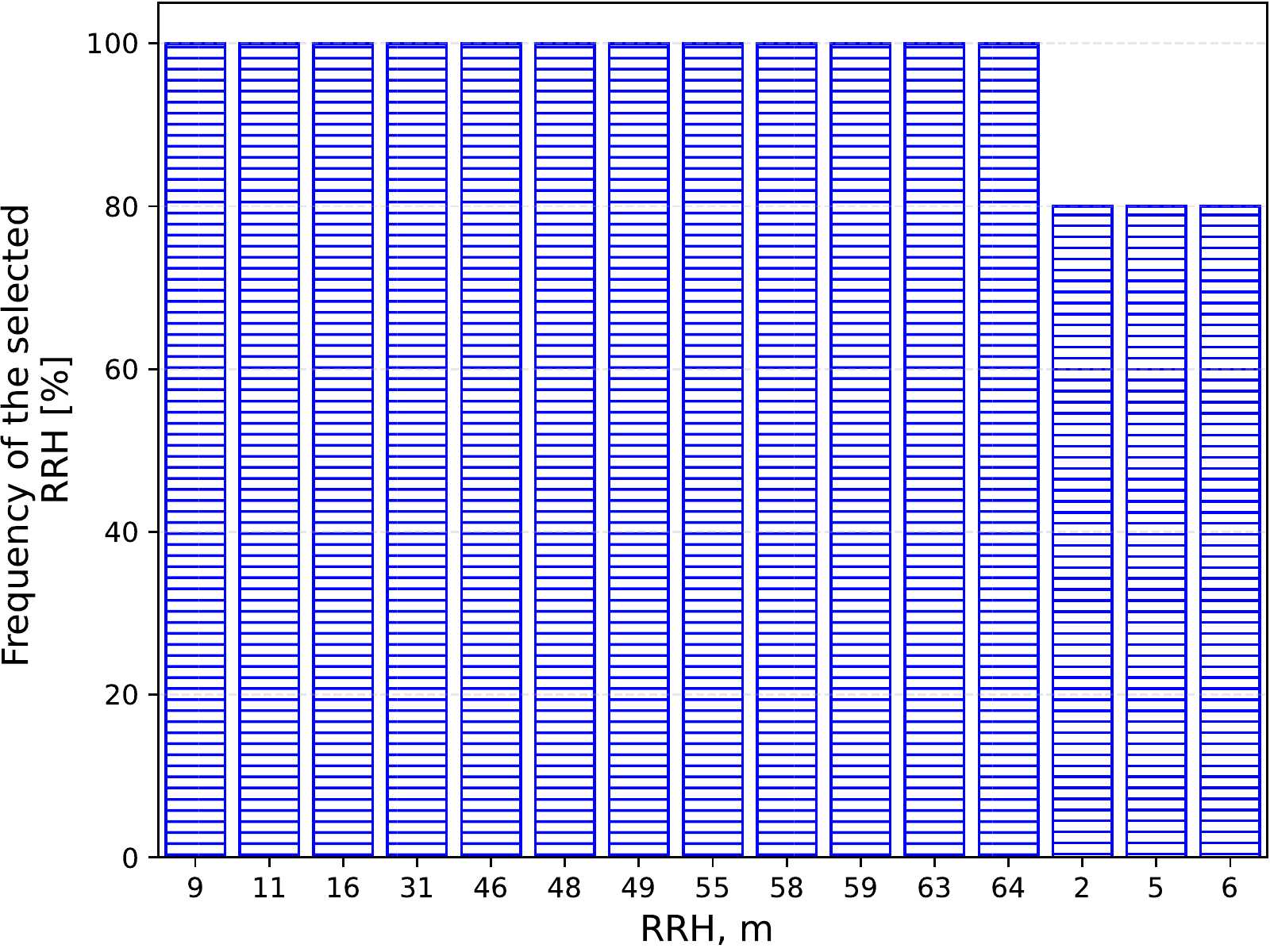}}
		\hfill
		\\
		\vspace{-0.0cm}
		\subfloat[$M'=15$ RRHs \label{fig:frequency_antenna_plot}]{%
			\includegraphics[width=0.85\linewidth]{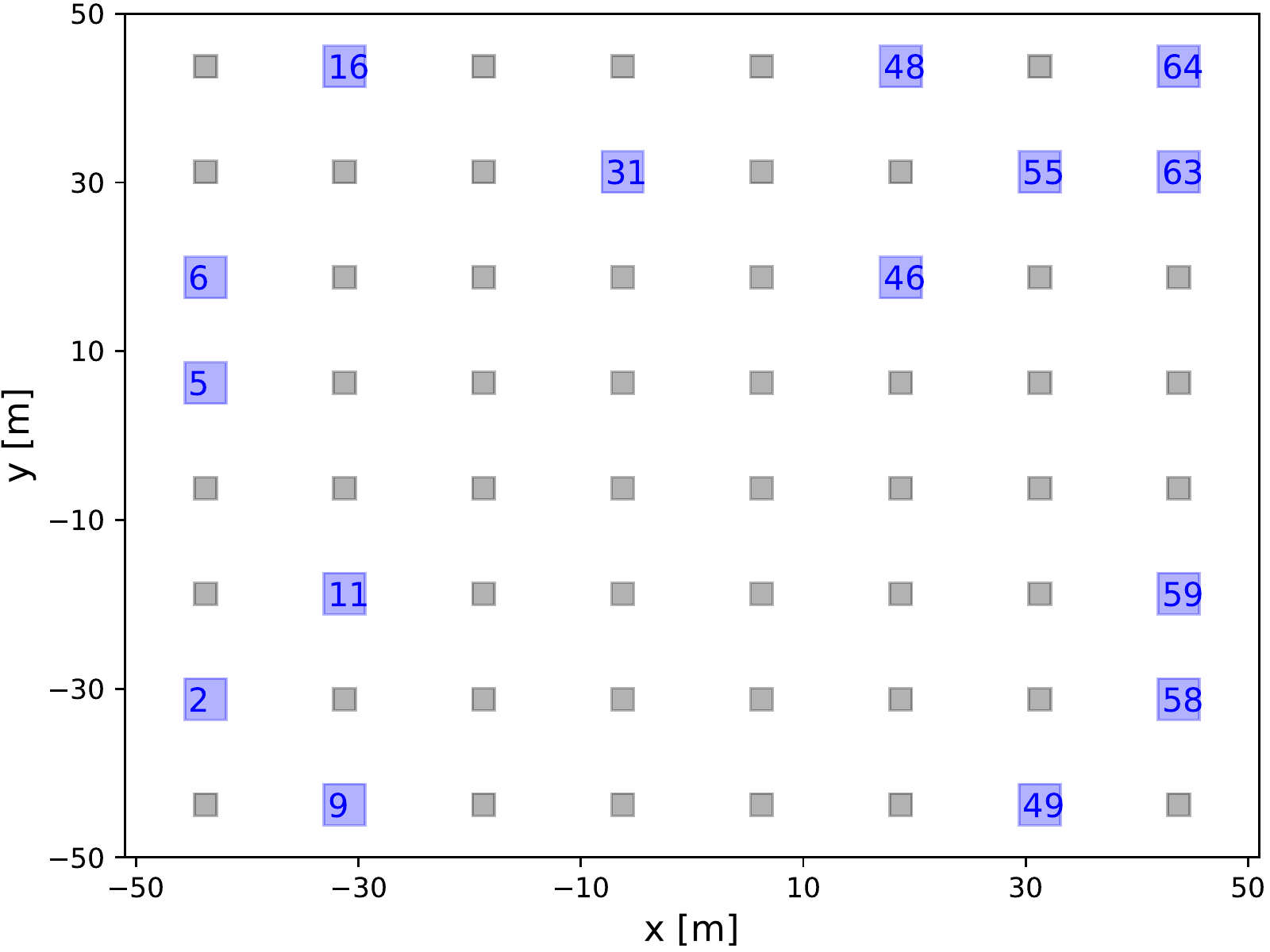}}
		\vspace{0.3cm}\caption{The most relevant RRHs in the ROI. (a) The occurrence of the most selected RRHs. (b) Actual positions of the $M' = 15$ RRHs.}
		\label{fig:Top_15_Selected_Antennas}\vspace{-0.8cm}
	\end{figure}

	An additional advantage of using the RRH selection strategy, is that it allows us to identify and possibly interpret the most relevant positions of the RRHs. This can be beneficial especially when it is hard to predict the propagation environment a priori, and thus no optimal configuration design of RRH positions can be foreseen. Fig. \ref{fig:Top_15_Selected_Antennas} depicts the $M' = 15$ most frequently selected RRHs among $M \in \{55,42,36,32,30\}$ and $\gamma = 3.0$. It can be noticed that the learning strategy consistently selects a number of RRHs and favors those in corners of the ROI. This suggests that certain positions of spatially distributed RRHs provide a more diverse view on the propagation environment and thus have a higher impact on minimizing the localization error during the selection process. \vspace{-0.3cm}
	\section{Conclusion}\label{conclusion}
	\vspace{0.0cm}
	In this work, we presented and investigated a learning-based strategy for RRH subset selection in a DAS for wireless localization. To alleviate the frontahul capacity constraints, we propose a scalable DNN-based model that incorporates a RRH selection layer and allows for end-to-end training over a discrete set of RRHs. We show that the selection strategy comes at the cost of small performance degradation. Further, the channel information obtained from the selected RRHs is utilized to estimate the  position of the transmitter and the inherent uncertainty in its estimates. A spatially distributed antenna system renders a more uniform localization error compared to centralized massive MIMO. Finally, the learned parameters in the RRH selection layer can be further probed to allow the analysis of position importance of the RRHs, possibly help the interpretation and facilitate design choices. 
	
	\vspace{-0.0cm}
	\small \section*{Acknowledgment}
	\vspace{0.00cm}
	This work has been funded by \"OBB Infrastruktur AG. The financial support by the Austrian Federal Ministry for Digital and Economic Affairs, the National Foundation for Research, Technology and Development and the Christian Doppler Research Association is gratefully acknowledged.
	\vspace{0.3cm}
	
	\bibliographystyle{IEEEtran}
	\bibliography{References}
	
	
	
\end{document}